\documentstyle[12pt,epsf]{article}
\newcommand {\Ljb} {\mbox{$L _ {\mathrm{JB}}$}}
\newcommand {\oalf} {\mbox{${\cal O}(\alpha)$}}
\newcommand {\Qt } {\mbox{$Q^2_{\tau }$}}
\newcommand {\xjb} {\mbox{$x_  {\mathrm{JB}}$}}
\newcommand {\yjb} {\mbox{$y_  {\mathrm{JB}}$}}
\newcommand {\qjb} {\mbox{$Q^2_{\mathrm{JB}}$}}
\newcommand {\Qz } {\mbox{$Q^2_{\tau}$}}

\newcommand {\LQZ} {\mbox{${\mr L}_{ \tau} $}}
\newcommand{\nn}{\noindent}

\newcommand{\bq}{\begin{equation}}
\newcommand{\eq}{\end{equation}}
\newcommand{\ba}{\begin{eqnarray}}
\newcommand{\ea}{\end{eqnarray}}

\newcommand {\oa} {\mbox{${\cal O}(\alpha)$}}
\newcommand{\mr}{\mathrm}
\newcommand{\nl}{ \nonumber \\}
\begin{document}
\thispagestyle{empty}
\onecolumn
\noindent
DESY 92--175
\\
December 1992
\\
Phys. Letters B301 (1993) 447
\\
revised July 1995

\vspace*{1.50cm}
{ \large \bf \begin{center}
 
LEPTONIC QED CORRECTIONS TO THE PROCESS   \\
$ep \longrightarrow eX$ IN JAQUET-BLONDEL VARIABLES
 
\vspace*{.5cm}
\end{center}
     } 

\begin{center}

\vspace*{1.3cm}
{\large \bf Arif Akhundov} \\
{\large \it Institute of Physics, Azerbaijan Academy of Sciences  \\
pr. Azizbekova 33, Baku 370143, Azerbaijan}
\\
 
\vspace*{.9cm}
{\large \bf   Dima Bardin, \, Lida Kalinovskaya} \\
{\large \it
Joint Institute for Nuclear Research, Dubna  \\
Head Post Office P.O. Box 79, Moscow 101000, Russia}
\\
\vspace*{.9cm}
{\large \bf    Tord Riemann} \\
{\large \it
DESY -- Institut f\"ur Hochenergiephysik
 \\} {\large \it
Platanenallee 6, O-1615 Zeuthen, Germany}
\end{center}
 
\vspace*{0.5cm}
\vfill
\thispagestyle{empty}
{\large
\centerline{\bf ABSTRACT}
\vspace*{.3cm}
\nn
\normalsize
For the study of deep inelastic scattering at HERA, the use of
Jaquet-Blondel variables is advantageous in several respects.
We calculate the complete leptonic
${\cal O}(\alpha)$
QED corrections for
the reaction $ep \rightarrow eX$                in these variables.
All but one phase space integrations are performed analytically.
After exponentiation of soft photon corrections, an  accuracy of
at least  1\% is matched.
Numerical results are presented
and compared to estimates based on the leading logarithmic approximation.

Equation (17) is corrected in accordance with equation (7.44) of DESY
94--115. The misprint did not influence the numerics.
} 
\vspace*{.5cm}
 
\bigskip
\vfill

\newpage
\section{Introduction}
At HERA, the deep-inelastic scattering of electrons off protons,
\ba
e(k_1) + p(p_1) \rightarrow e(k_2) + X(p_2),
\label{born}
\ea
allows the study both of the basic interactions of
electrons and protons and of the proton structure
                in a completely new kinematic
region~\cite{heraws}.
Reaction~(\ref{born}) has to be analyzed together with the corresponding
photonic corrections from the radiative process,
\ba
e(k_1) + p(p_1) \rightarrow e(k_2) + X(p_2) + n\gamma(k).
\label{brems}
\ea
Bremsstrahlung may substancially contribute to the
observed net cross section. Its amount crucially
       depends both
 on the         experimental cuts applied to   the photonic phase space and
on the kinematic variables and their allowed regions of variation.
During a longer period, the use of    leptonic variables dominated in the
literature.
At HERA,
these have certain disadvantages:
QED corrections may become extremely large,
and in certain kinematic regions it is impossible to separate energetically
the final electron from the bremsstrahlung photon.
More advantageous are the hadronic variables  on which
the structure functions depend. Unfortunately, their experimental
determination is also nontrivial or even impossible.
A natural compromise has been proposed with the
{\em Jaquet--Blondel} (JB) variables~\cite{jablo}.
During the 1991 HERA Workshop~\cite{structure}, 
features of the various sets of variables have been discussed
in detail.
Here we will concentrate on the case of JB variables.
These variables are not influenced neither by the leptonic nor
the longitudinal hadronic degrees of freedom:
\ba
 \qjb = \frac{(\sum_h {\vec{p}}_{h\perp})^2}{1-\yjb}
     = \frac {
({\vec p}_{2\perp})^2
} {1-\yjb} ,\qquad
 \yjb = y_h = \frac{-2 p_1 Q_h}{S}, \qquad
\xjb =\frac{\qjb}{ \yjb S},
\label{xjb}
\ea
with $S=-(k_1+p_1)^2$ and $Q_h=p_2-p_1$.
Using the JB variables, one has the opportunity
to search for a phase space parametrization with
an early integration over the leptonic degrees of freedom.
In another approach, we did not use this opportunity since we tried there
to handle leptonic, mixed, and
hadronic variables in a common formalism as much as
possible~\cite{teupitz,5a}.
For details of the derivation of the formulae which will be presented
here, we refer to that article.
 
In this letter,
we take advantage of the above-mentioned properties of the   JB variables.
We have performed all but one
                                  phase space integrations for the
                                  double-differential
cross section.
We explicitely
scetch the rather compact complete
${\cal O}(\alpha)$
{\em leptonic}
corrections from reaction~(\ref{brems})
 to  the double-differential cross section of
reaction~(\ref{born}), including soft
photon exponentiation.
The leptonic corrections arise from the
radiation of photons from the leptonic
part of the Feynman diagrams.
They are much larger than the other ones --
quark- and lepton-quark interference bremsstrahlung.
Further, leptonic bremsstrahlung may be treated {\em model--independently},
relying on some set of structure functions which need not necessarily
be related to the quark-parton model.
 
In Section~2, the kinematics is explained and the basic formulae
are collected. Section~3 contains numerical results and their
discussion.

\section{Kinematics and phase space integrations}
\subsection{The Born cross section}
The cross section of~(\ref{born}) is
\ba
\frac {d^2 \sigma_{\mr B}} {  dQ^2  dy   }
  = \frac{2 \pi \alpha^{2}}{S Q^4         } \,
    \sum_{i=1}^3 A_i(x  ,Q^2  )
    {\cal{S}}_{i}^{\mr B} (Q^2 ,y   ),
\label{eqBorn}
\ea
where,
of course, for the lowest-order cross section  the different sets of kinematic
variables agree; e.g. $\qjb=Q^2_h$, $\yjb=y_h$, $\xjb=x_h$.
Further,
\ba
{\cal S}_{1}^{\mr B} &=& Q^2,  \nl
{\cal S}_{2}^{\mr B} &=& 2    (1- y  ) S^2        ,\nl
{\cal S}_{3}^{\mr B} &=& 2(2-y)Q^2  S        .
\label{eqS3B}
\ea
The $A_i           $ are proportional to the
structure functions ${\cal F}_{1,2,3}^{\mathrm{NC}}(x_h,Q^2_h) $
             describing the electroweak
interactions of leptons and nucleons; their arguments have to be defined
by the hadronic kinematics:
\begin{eqnarray}
A_{1}             &\equiv& {(2MW_{1})} =
{\cal F}_{1}^{\mathrm{NC}}(x_h,Q^2_h)\;,  \\
A_{2}  &\equiv&
\frac{1}{y_h  S}
 ( \nu W_{2})  = \frac{ 1}     { y_h S}
{\cal F}_{2}^{\mathrm{NC}}(x_h,Q^2_h)\;, \nonumber \\
A_{3} &\equiv& {1\over 2y_h  S}{({\nu}W_{3})} = { 1 \over 2y_h  S }
{\cal F}_{3}^{\mathrm{NC}}(x_h,Q^2_h)\;.\nonumber
\end{eqnarray}
The    structure functions are
defined as usually:
\begin{eqnarray}
{\cal F}_{1,2}^{\mathrm{NC}}(x,Q^2) &=&
F_{1,2}(x,Q^2) + 2 |Q_{l}|v_{l} \chi G_{1,2}(x,Q^2)     +
\chi^{2}( v_{l}^{2} + a_{l}^{2} ) H_{1,2}(x,Q^2),  \nonumber \\
\pagebreak[4]
{\cal F}_3^{\mathrm{NC}}(x,Q^2) &=&
  -2 \chi Q_{l}a_{l} G_{3}(x,Q^2)
      - 2\chi^2 Q_{l} v_{l} a_l H_{3}(x,Q^2),
\label{f123}
\end{eqnarray}
where $Q_l$, $v_l$ and $a_l$ are the corresponding charge
$(Q_{l^{\pm}}=\pm 1)$,
vector and axial-vector couplings
of the lepton with the $Z$ boson:
$ v_{l}=1-4|Q_{l}| \sin^{2}\theta_{w},\;\;a_{l}=1$,
and $\theta_{w}$ is the weak mixing angle.
Further,
\ba
\chi \equiv \chi(Q^2) =
{G_\mu \over\sqrt{2}}  {M_{Z}^{2} \over{8\pi\alpha}}{Q^2 \over
{Q^2+M_{Z}^{2}}}.
\label{chi}
\ea
The $F,G,H$ are structure functions
    parametrizing the general hadronic tensor for
the $|\gamma|^2, {\gamma}Z$, and $|Z|^2$ contributions to the Born
cross-section, respectively.
 
For the radiative process, the sum at the right-hand side
in~(\ref{eqBorn}) will become the integrand of a three-dimensional
phase space integral. Further, the kinematic weights ${\cal S}_i$
which accompany the structure functions become complicated
expressions of the invariants.
The specific problem which has to
   be solved for each set of variables separately is the subsequent
analytic integration.
In the next chapter, we comment on that in terms of  Jaquet--Blondel (JB)
 variables and perform
two integrations explicitely. Then,       we are left with a
one-dimensional integral which has to be performed numerically.
 
\subsection{The phase space integration                    }
With JB~variables, the following parametrization of the phase
space may be derived~\footnote{
Whenever necessary, we retain the electron mass $m$ although
the ultra-relativistic limit is assumed.}~\cite{5a}:
\ba
\Gamma = \frac{\pi^2} {4  }   
\int d\qjb  \,  d\yjb
\int_{m^2}^{\tau_{\max}} d\tau \frac{1}{4 \pi} \frac{\tau - m^2}{\tau}
\int_{-1}^1 {d\cos{\vartheta_R} \int_0^{2 \pi} d{\varphi_R}} \int d\Gamma_h.
\label{gmto2}
\ea
Here, the
 \ba
 d\Gamma_h = \prod_i \frac{d\vec{p_i}}{2p_i^0}
             \delta^{(4)}(p_2-\sum_ip_i)
 \label{dgm}
 \ea
is the phase space element of the final hadron system
which will be completely absorbed into the definition of the
hadronic structure functions.
Further, we use
the   invariant mass $\tau=-(k_2+k)^2$
of the ($e,\gamma$)
compound and the photonic angles
$\vartheta         _R$,
$\varphi         _R$
in the rest system of this compound,  defined by $\vec{k_2} + \vec{k} = 0$.
Further, $m$ is the electron mass, and the upper limit of the $\tau$ variation
is
\ba
\tau_{\max} =
(1-\xjb)(1-\yjb) S.
\ea
In these variables, the double-differential cross section becomes:
\ba
\frac {d^2 \sigma_{\mr R}} { d\qjb d\yjb }
       = \frac{2 \alpha^{3}}{S      }
         \int\! d\tau
\sum_{i=1}^3
\frac{1}{Q_h^4}
A_i(x_h,Q^2_h)
{\cal S}_{i}(\qjb,\yjb,\tau),
\label{dsigR}
\ea
\ba
 {\cal S }_i
 (\qjb,\yjb,\tau)
 =\frac{1}{4 \pi} \frac{\tau-m^2}{\tau}
 \int {d\cos{\vartheta_R}}    {d{\varphi_R}} \;
 {\cal S}_i(\qjb,\yjb,\tau,\cos{\vartheta_R},\varphi_R ).
\label{eq313}
\ea
The functions $ {\cal S}_i(\qjb,\yjb,\tau,\cos{\vartheta_R},\varphi_R )$
are calculated from the squared sum of the Feynman diagrams.
Their explicit expressions may be found elsewhere~\cite{xxx,5a}.
In the derivation of~(\ref{dsigR}) we took advantage of the fact
that the hadronic variables, and thus the hadronic structure functions,
are independent of the photonic angles.
The hadronic transferred momentum squared $Q_h^2$ and $\qjb$
are related by the
following relation containing the additional variable $\tau$:
\ba
  Q^2_h  \equiv  (p_2-p_1)^2 =
\qjb + \frac{\yjb}{1-\yjb} (\tau - m^2).
\label{qhqjb}
\ea
For Born kinematics, $\tau$ approaches its lower limit
and $Q_h^2$ and $\qjb$ agree.
The integral over $\tau$ in~(\ref{dsigR})
has to be performed numerically as long as one cannot neglect
the difference between $\qjb$ and $Q_h^2$ --    the structure functions
are depending on $Q_h^2$ and thus also on $\tau$.
At the other hand,
the integral over the photonic angles may be performed analytically.
For details we refer to~\cite{777,5a}
where the integration techniques are explained and most of the
 integrals used may be found.
Again, we quote here only the final result which has been obtained by
means of the program for analytical calculations
{\tt SCHOONSCHIP}~\cite{schoonschip}:
\ba
{\cal S}_1(\qjb,\yjb,\tau)   =
    \qjb
    \left[ \frac{1}{z_2} \left(\LQZ -2\right)
 +  \frac{z_2 }{4 \tau^2} \right]
 +  \frac{ 1 - 8 \yjb }{4(1-\yjb)}  
 +  \LQZ \left( \frac{z_2}{2 \Qt }+\frac{\yjb}{1-\yjb}\right),
\label{S1}
\ea
\ba
{\cal S}_2(\qjb,\yjb,\tau) & = &
      S^2 \Biggl\{  2(1-\yjb)
    \left[ \frac{1}{z_2} \left( \LQZ - 2 \right)
  + \frac{z_2}{4\tau^2} \right]   
  - \frac{1}{\Qt} \left[ 2( \LQZ -2)
  +  \frac{1}{2}(1-{\yjb}^2) \right] \nl
 &+&\frac{\qjb} {\mbox{$Q^4_{\mr  \tau} $}} (1-\yjb)
    \left[ 1 -  (1+\yjb)(\LQZ -3) \right]
  - \frac{\mbox{$Q^4_{\mr {JB}}$}}
         {\mbox{$Q^6_{\mr \tau}$}} (1-\yjb)^2 (\LQZ-3)
                     \Biggr\},
\label{S2}
\ea
\ba
{\cal S}_3(\qjb,\yjb,\tau) & = &
      S \Biggl\{ 2 \qjb (2-\yjb)
      \left[ \frac{1}{z_2} (\LQZ - 2)
    + \frac{z_2}{4\tau^2} \right]             
 +  \frac{\yjb( 1 + \yjb^2 )} { 1-\yjb } \LQZ +5   \nl
& &-~ \frac{ 7 \yjb }{2(1-\yjb)}-(1-\yjb)(5+2\yjb)  \nl
& &+~ \frac{\qjb}{\Qt} (1-\yjb) \left[(1-\yjb)
                              \left( 3-\frac{2 \qjb }{ \Qt }\right)
  (\LQZ - 2)
  + 12-5 \LQZ \right]
                     \Biggr\},
\label{S3}
\ea
where
\ba
 z_2 & =&-2 k_2 k = \tau - m^2, \\
\LQZ &= & \mbox{ln} {\frac{\Qt} {m^2 \tau}}, \\
\Qt  &= & \qjb + \frac{\tau - m^2 }{1-\yjb}.
\ea
\subsection{The infra-red problem and  soft photon exponentiation}
 The integral (\ref{dsigR}) diverges at $z_2=0 $ ($ k=0 $).
This may be seen from~(\ref{S1}--\ref{S3});
the kinematic factors ${\cal S}_i$
 contain the Born functions~(\ref{eqS3B}) together with the common, divergent
factor $(\LQZ - 2)/z_2$.
The infra-red divergent part of~(\ref{dsigR})
may be isolated:
\ba
\frac {d^2 \sigma_{\mr R}} {d\qjb d\yjb} =
\left[
\frac {d^2 \sigma_{\mr R}} {d\qjb d\yjb} -
\frac {d^2 \sigma_{\mr R}^{\mr {IR}}}{d\qjb d\yjb}
\right]
+
\frac {d^2 \sigma_{\mr R}^{\mr {IR}}}{d\qjb d\yjb} \equiv
\frac {d^2 \sigma_{\mr R}^{\mr F }}{d\qjb d\yjb} +
\frac {d^2 \sigma_{\mr R}^{\mr {IR}}}{d\qjb d\yjb}.
\ea
Here
$ {d^2 \sigma_{\mr R}^{\mr F}}  / {d\qjb d\yjb} $
is finite at $ k\rightarrow 0 $;
it is defined as follows:
\ba
\frac {d^2 \sigma_{\mr R}^{\mr F}} { d\qjb d\yjb }
      &=&
\frac{2 \alpha^{3}}{S      }
         \int\! d\tau
\sum_{i=1}^3
\Bigl[
\frac{1}{Q_h^4}A_i(x_h,Q^2_h){\cal S}_{i}(\qjb,\yjb,\tau)
\nl
&-&
\frac{1}{Q_{\mr {JB}} ^4}  A_i(\xjb,\qjb){\cal S}_i^{\mr B} (\qjb,\yjb )
F^{\mr {IR}}(\qjb,\yjb,\tau)
\Bigr]
,
\label{SRF}
\ea
\ba
F^{\mr {IR}}
(\qjb,\yjb,\tau),
=
     \frac{1}{z_2} \left(\LQZ             -2 \right)
  - \frac{1}{(1-\yjb)\Qz  } \LQZ + \frac{1}{\tau}.
\label{FIR}
\ea
From the above equations, it is easy to derive an explicit
expression for
the isolated infra-red divergent part $d^2\sigma_{\mr R}^{\mr {IR}}$.
It contains an integral over $\tau$ which may be performed
explicitely using dimensional regularization~\cite{dimreg,5a}.

Finally, the resulting cross section is
\ba
\frac {d^2 \sigma        } {d\qjb d\yjb} =
\frac {d^2 \sigma_{\mr R}^{\mr F }}{d\qjb d\yjb}
+
\frac {d^2 \sigma_{\mr B}} {d\qjb d\yjb}
\left[
1 + \frac{\alpha}{\pi} \delta^{\mr {VR}}
(\qjb,\yjb)
\right],
\ea
\ba
\delta^{\mr {VR}}
(\qjb,\yjb)
\equiv
\delta_{\mr {vert}}
(\qjb,\yjb)        +
\delta_{\mr R}^{\mr {IR}}
(\qjb,\yjb).
\ea
As mentioned, from~(\ref{SRF}) one derives
\ba
\delta_{\mr R}^{\mr {IR}}
(\qjb,\yjb)
=
\int d\tau F^{\mr {IR}}
(\qjb,\yjb,\tau),
\ea
with $F^{\mr {IR}}$ defined in~(\ref{FIR}).
The QED vertex correction
$\delta_{\mr {vert}}$ compensates for the infra-red divergence
 ${\cal P}^{\mr {IR}}(\mu )$
of the soft
bremssstrahlung contribution,
\ba
\delta_{\mr {vert}}
(\qjb,\yjb)
  = - 2
  {\cal P}^{\mr {IR}}(\mu )
 \left( \Ljb                       -1 \right)
 -   \frac{1}{2} \Ljb^2
 +
 \frac{3}{2} \Ljb
 +  {\mr {Li}}_2(1) - 2,
\label{vert}
\ea
where
\ba
\Ljb = \ln \frac{\qjb  }{m^2},
\ea
and ${\mr {Li}}_2(1)$ is the Euler dilogarithm.
The net photonic correction $\delta^{\mr {VR}}$ is finite:
\ba
\delta^{\mr {VR}}(\qjb,\yjb)
&=&  \delta^{\mr{inf}}(\qjb,\yjb)
-\mbox{ln}^2X_{\mr {JB}}
+
\left[\mbox{ln}(1-\yjb)-1\right]
\mbox{ln}X_{\mr {JB}}
\nl
&-&\frac{1}{2}
 \mbox{ln}^2 \left[ \frac{(1 - \xjb)(1 - \yjb)}{\xjb\yjb} \right] \nl
&+& \frac{3}{2}\Ljb
-   {\mr {Li}}_2 (X_{\mr {JB}})
- 2 {\mr {Li}}_2 (1-{X_{\mr {JB}}}) - 1 ,
 \label{sumVR}
\ea
where
\ba
\delta^{\mr {inf}} (\qjb,\yjb)          =
\left( \Ljb - 1 \right)
\mbox{ln} \frac{ (1-\xjb)(1-\yjb)}{1 - \xjb(1-\yjb)},
\ea
\ba
 X_{\mr {JB}} = \frac{1}{\xjb\yjb} \left[ 1 - \xjb(1-\yjb) \right].
\ea
 
Finally,
the effect of multiple soft photon emission may be taken into account
by soft photon exponentiation:
\ba
\frac{\alpha}{\pi} \delta^{\mr {VR}}(\qjb,\yjb)
&\rightarrow&
\frac{\alpha}{\pi} \delta^{\mr {VR}}_{\mr {exp}}(\qjb,\yjb)
\nl
&=&
\frac{\alpha}{\pi} \left[ \delta^{\mr {VR}}(\qjb,\yjb)
-\delta^{\mr {inf}}(\qjb,\yjb)  \right]
+
\left\{
\mbox{exp} \{ \frac{\alpha}{\pi} \delta^{\mr {inf}}(\qjb,\yjb) \} - 1
\right\}.  \nl
\ea
 
\section{Results and discussion}
The numerical results are obtained with an extended version of the
FORTRAN program {\tt TERAD91}~\cite{5a}.
We use exactly the same specifications of kinematics, structure
functions, weak Standard Model parameters for the $Z$~amplitude, etc.
as have been used at the 1992 HERA
workshop~\cite{herarc}. For this reason,
    the running of the QED coupling has not been taken into account
in the matrix elements.
We use first the ${\cal O}(\alpha)$ approximation.
The table~1
contains the QED corrections $\delta$, which are defined as follows:
\ba
\delta = \left[
\frac
{
   d^2 \sigma  /                  {d\qjb d\yjb}
}
{
 { d^2 \sigma_{\mr B} }  /        {d\qjb d\yjb}
}
-1
\right]\times 100\%.
\ea
 
\begin{table}[hbt]
\label{tab3}
\begin{center}
\begin{tabular}{|c|c|c|c|}
\hline
       &      &  TERAD                 &  HELIOS      \\
$\xjb$ &$\yjb$&   complete             &  LLA         \\
       &      &  $ \oa $               &  $ \oa $     \\
\hline
\hline
$10^{-3}$&$0.01$&  $-0.19$     & $-0.01$  \\
\hline
         &$0.10$&  $-1.17$     & $-1.1 $   \\
\hline
         &$0.50$&  $-4.59$     & $-4.5 $   \\
\hline
         &$0.90$&  $-12.11$    & $-11.7$   \\
\hline
         &$0.99$&  $-24.15$    & $-22.2$   \\
\hline
\hline
$10^{-2}$&$0.01$&  $-0.22$     & $-0.2$   \\
\hline
         &$0.10$&  $-1.49$     & $-1.4$   \\
\hline
         &$0.50$&  $-5.50$     & $-5.3$   \\
\hline
         &$0.90$&  $-13.98$    & $-13.5$  \\
\hline
         &$0.99$&  $-27.31$    & $-25.2$  \\
\hline
\hline
$0.10$   &$0.01$&  $-0.29$     & $-0.3$   \\
\hline
         &$0.10$&  $-1.82$     & $-1.8$   \\
\hline
         &$0.50$&  $-6.19$     & $-6.0$   \\
\hline
         &$0.90$&  $-15.86$     & $-15.2$  \\
\hline
         &$0.99$&  $-30.57$    & $-28.4$  \\
\hline
\hline
$0.50$   &$0.01$&  $-0.58$     & $-0.5$   \\
\hline
         &$0.10$&  $-3.29$     & $-3.1$   \\
\hline
         &$0.50$&  $-11.11$     & $-10.7$  \\
\hline
         &$0.90$&  $-24.17$    & $-22.8$  \\
\hline
         &$0.99$&  $-40.75$    & $-37.2$  \\
\hline
\end{tabular}
\caption[]
{ \it
Leptonic radiative corrections $\delta$ of       \oalf in percent;
first column  this calculation, second column a LLA
calculation~\cite{mixjb}.
}
\end{center}
\end{table}
 
The corrections are small for small
$\xjb$ and $\yjb$.
With rising $\yjb$, they may become large. In magnitude, they fall
in between the leptonic and the considerably smaller hadronic
corrections.

In the table, we also show numerical results
from another calculation, which is restricted to the leading
logarithmic approximation~(LLA).
These numbers have been taken from table~3
in~\cite{herarc} and are based on formulae published in~\cite{mixjb}.
The agreement is as good as one can expect for a LLA calculation.
It is much better than has been observed for the case of
leptonic variables, where certain kinematic singularities which are not related
to the large logarithms from the particle masses, are much stronger pronounced.
A naive conclusion from the comparison could be that experimentalists
who apply radiative corrections to data should restrict
                 to the use of the LLA formulae, thus optimizing the
numerical computings.
                   Let us remind here that this
would not be too advantageous since both the exact \oalf\
and the approximated formulae are one-dimensional integrals.

\begin{figure}[thb]
\label{f1}
\begin{center}
{\hspace*{+1.0cm}
\mbox{
\epsfysize=16.cm
\epsffile[0 0 530 530]{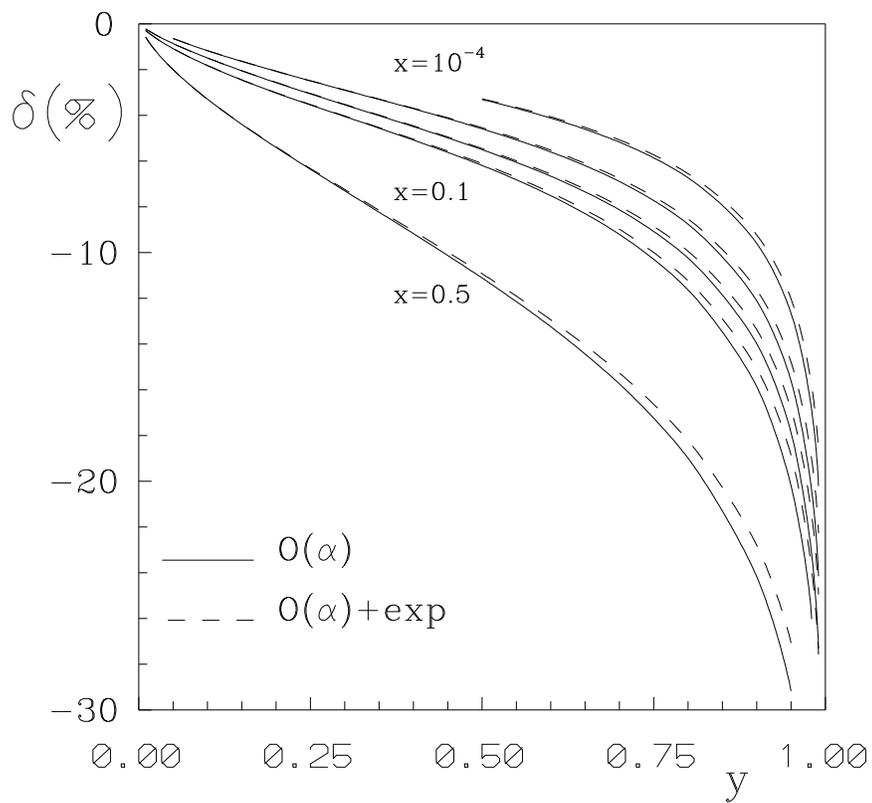}
}
}
\end{center}
\caption[]
{\it
Leptonic radiative corrections $\delta$ of       \oalf\
without (solid curves) and
with (broken curves) soft photon exponentiation in JB~variables;
parameter: $\xjb = 0.0001, 0.001, 0.01, 0.1, 0.5$.
}
\end{figure}
 
In figure~1,
we show the same corrections, but now including soft photon
exponentiation (SPE; broken curves).
Its inclusion effects the cross sections only there, where the
corrections are numerically large, i.e. near $y=1$.
 
{\em To summarize}, in this letter the complete \oalf\ QED corrections,
including soft photon exponentiation,  to deep inelastic $ep$ scattering
are obtained in JB variables for the first time in form of a
semi-analytical expression.
The numerical improvement compared to a LLA~calculation, which also
contains a one-dimensional integration, is minor.

\section{Acknowledgements}
Three of us (A.A, D.B., L.K.) would like to thank the DESY~--~Institut
f\"ur Hochenergiephysik for the kind hospitality at Zeuthen where
the major activity for the project took place. Thanks also to
J.~Bl\"umlein for discussions.
 

\end{document}